\documentclass[prl,twocolumn,showpacs,preprintnumbers,amsmath,amssymb,apsrev]{revtex4}

\usepackage{graphicx}
\usepackage{dcolumn}
\usepackage{bm}
\usepackage{subfigure}
\usepackage[sort&compress]{natbib}

\begin{document}

\title{Quantitative LEED I-V and \emph{ab initio} study of the Si(111)-3x2-Sm surface structure
and the missing half order spots in the 3$\times$1 diffraction
pattern.}
\author{C. Eames, M. I. J. Probert, S. P. Tear}
\email[Corresponding author. E-mail:\,\,\,]{spt1@york.ac.uk}
\affiliation{Department of Physics, University of York, York YO10
5DD, United Kingdom}

\begin{abstract}
We have used Low Energy Electron Diffraction (LEED) I-V analysis
and \emph{ab initio} calculations to quantitatively determine the
honeycomb chain model structure for the Si(111)-3$\times$2-Sm
surface. This structure and a similar 3$\times$1 recontruction
have been observed for many Alkali-Earth and Rare-Earth metals on
the Si(111) surface. Our \emph{ab initio} calculations show that
there are two almost degenerate sites for the Sm atom in the unit
cell and the LEED I-V analysis reveals that an admixture of the
two in a ratio that slightly favours the site with the lower
energy is the best match to experiment. We show that the I-V
curves are insensitive to the presence of the Sm atom and that
this results in a very low intensity for the half order spots
which might explain the appearance of a 3$\times$1 LEED pattern
produced by all of the structures with a 3$\times$2 unit cell.
\end{abstract}

\pacs{61.46.-w, 61.14.Hg, 68.43.Bc}

\maketitle

\section{I. Introduction}

\noindent The prospect of creating an ordered one dimensional
system has lead to the extensive study of chain structures grown
on surfaces. The alkali metals (AM) form such a chain structure as
part of a 3$\times$1 reconstruction on the Si(111) surface with an
AM coverage of 1/3 ML (Ref. \cite{AM1,AM2} and therein). At a
coverage of 1/6 ML the alkali earth metals (AEM) and the rare
earth metals (REM) form a 3$\times$2 reconstruction (Ref.
\cite{AEM1,AEM2,REM1,REM2} and therein). There is a wealth of
experimental evidence from STM, LEED and spectroscopic techniques
to suggest that in these 3$\times$ structures there is a common
structure for the reconstructed silicon (Ref.
\cite{AEM1,AEM2,REM1,REM2,3xn1,3xn2,3xn3,3xn4,3xn5} and therein).
The honeycomb-chain channel model (HCC) is now regarded by many as
the most plausible of the candidate structures \cite{HCC1, HCC2,
HCC3}. In the HCC model there are parallel ordered one dimensional
lines of metal atoms sited in a silicon free channel. These are
separated by almost flat honeycomb layers of silicon.

The 3$\times$1 system has been studied using LEED I-V analysis
with Ag, Li and Na as the deposited metal atoms \cite{3xn5}.
Similar I-V curves were obtained in each case and the authors
conclude that a common reconstruction of silicon atoms is
responsible for the LEED I-V curves, which are insensitive to the
presence of the metal atom. However, the authors did not attempt a
structural fit. The LEED pattern for the 3$\times$2 surfaces
exhibits odd behaviour in that it indicates a 3$\times$1
periodicity. Many workers have suggested that disorder in the
position of the metal atom is the cause. A Fourier analysis of a
random tesselation of a large sample of registry shifted
3$\times$2 unit cells has been carried out by Sch\"{a}fer et al.
\cite{3xn6}. They show that this simulation of long range disorder
in the position of the metal atom does produce a 3$\times$1
periodicity in reciprocal space. Alternatively, Over et al.
\cite{3xn7} have suggested that the substrate and silicon adatoms
could be acting as the dominant scattering unit with the metal
atoms sitting in `open sites'.

STM investigations of the 3$\times$2 and 3$\times$1 systems have
not provided much evidence of long range disorder in the location
of the metal atom apart from registry shifts introduced by a
coexisting c(6$\times$2) phase. Of particular relevance to this
work is the study of the Si(111)-3$\times$2-Sm system using STM
and an \textit{ab initio} calculation, carried out by Palmino et
al. \cite{REM1}. They have used the bias voltage dependence of the
STM images of the surface to isolate the features associated with
the honeycomb chain and the samarium atom and separate comparison
of these with simulated STM images obtained from the \textit{ab
initio} calculation show that the HCC structure is in good lateral
qualitative agreement with experiment.

In this study we have used LEED I-V analysis and several
\textit{ab initio} calculations to quantitatively investigate the
3$\times$2 reconstruction of the Si(111)-3$\times$2-Sm surface. We
show that the HCC structure gives good agreement with experiment.
We consider two HCC unit cells in which the samarium atom is
located in the T4 site or the H3 site. Palmino et al. \cite{REM1}
have found the energy difference of these two configurations to be
0.07 eV/Sm. We have calculated the atomic positions and the
energies of these two reconstructions and obtained LEED I-V curves
for this system and we show that a linear combination of the two
HCC structures is the optimum match to experiment with a ratio
that slightly favours the structure with the lower energy of the
two.

We have also used LEED I-V analysis to investigate the missing
half order spots for the 3$\times$2 unit cell. We show using
calculated I-V curves that for an individual unit cell the
intensity of the half order spots is significantly lower than that
of the spots that are visible in the experiments. We also show
that the calculated I-V curves do not differ significantly if the
samarium atom is not present. We offer this as evidence that
disorder over multiple unit cells is not needed to explain the
STM/LEED discrepancy for the 3$\times$2 systems and we suggest
that the order in the one dimensional chain may persist over large
length scales.

\section{II. Experimental}

\noindent A dedicated LEED chamber of in-house design
\cite{Carvalho} operating at a typical UHV base pressure of $\sim
10^{-10}$ mbar was used to carry out our experiments. The silicon
substrate was cleaned by flashing to $\approx1200^{\,\circ}$C
using an electron beam heater and then the sample was slowly
cooled through the $\approx900-700^{\,\circ}$C region over a
period of 15 minutes. A sharp 7$\times$7 LEED pattern resulted,
confirming that a clean Si(111)-7$\times$7 surface had been made.
Temperatures were monitored using an infra-red pyrometer.

\begin{figure}[h]
\centering
  \includegraphics[width=3.5in]{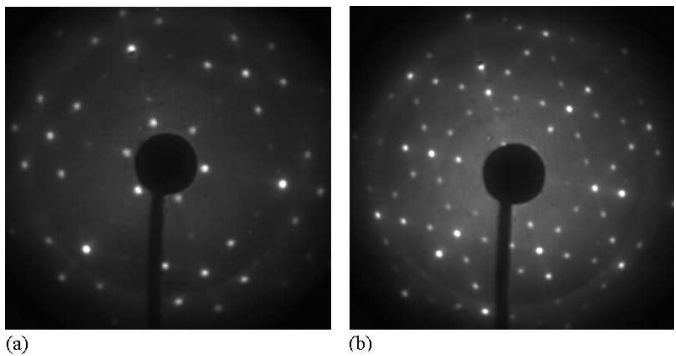}\\
  \caption{Experimental 3$\times$1 LEED spot pattern for the Si(111)-3$\times$2-Sm
  surface shown at (a) 40 eV and (b) 80 eV.}\label{fig:fig01}
\end{figure}

In the literature other workers \cite{Palmino, 3xn3, REM1} have
formed the Si(111)-3$\times$2-Sm structure by depositing 1/6 ML
onto a sample held at a temperature of $400-850^{\,\circ}$C
followed by annealing at this temperature for 20 min. In this work
the sample was prepared by depositing 1 ML of Sm from a quartz
crystal calibrated evaporation source onto the clean
Si(111)-7$\times$7 surface which was not at elevated temperature.
This was followed by a hot anneal at $\approx700^{\,\circ}$C for
15 minutes. A sharp 3$\times$1 LEED pattern was observed and
images of this are shown in figure \ref{fig:fig01}. Other workers
have observed some streaking in the 3$\times$1 LEED pattern that
is indicative of one dimensional disorder. We have not observed
such streaking in our diffraction patterns and we attribute this
to our preparation procedure. There is some variability in the
annealing temperature that can be used and temperatures in the
range $\approx700-900^{\,\circ}$C all gave a sharp diffraction
pattern. It is at around $1000^{\,\circ}$C that the pattern begins
to degrade as the samarium desorbs.

Images of the diffraction pattern were acquired over a 40-250 eV
range of primary electron energies in steps of 2 eV using a CCD
camera and stored on an instrument dedicated computer. For each
spot in the LEED pattern the variation in its intensity with
primary electron energy was recorded which resulted in a set of 42
I-V curves.

\begin{figure}[h]
\centering
  \includegraphics[width=3.3in]{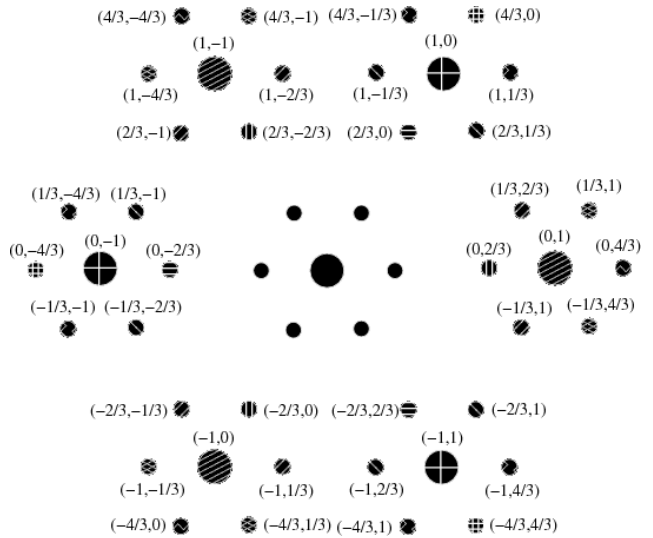}\\
  \caption{Labelled spots in the 3$\times$1 LEED pattern produced by the Si(111)-3$\times$2-Sm
  surface as it appears at 40 eV. The degeneracies of the spots are indicated by the pattern used to fill each
  spot.}\label{fig:fig02}
\end{figure}

Degenerate beams were averaged together to reduce the signal to
noise ratio and also to reduce any small errors that may have
occurred in setting up normal beam incidence. Figure
\ref{fig:fig02} defines the spot labelling system and the
degenerate beams. The experiment was repeated several times and
the I-V curves obtained during different experiments were compared
using the Pendry R-factor \cite{Rp}. The R-factor for I-V curves
obtained on different days was typically 0.1 or less which
indicates that the surface is repeatedly preparable. To further
reduce noise the I-V curves from separate experiments were
averaged together and a three point smooth was applied.

This set of 13 averaged I-V curves was used to fingerprint the
surface structure and allow comparison with the I-V curves
calculated for the various trial structures.

\section{III. \textit{Ab initio} Calculations}

\textit{Ab initio} calculations were performed using the
\texttt{CASTEP} code \cite{CASTEP}. The code was run on 30
processors in a parallel computing environment at the HPCx High
Performance Computing facility located at the CCLRC Daresbury
laboratory in the UK. We have geometry optimised two different
unit cells for the HCC structure (see figure \ref{fig:fig05} for
details of these). In the first unit cell the samarium atom is
located in the T4 site with respect to the first bulk-like silicon
layer and in the other unit cell the samarium atom is situated in
a H3 site. We will refer to the two structures as `T4' and `H3'.
The initial atomic positions were those that were obtained in the
\textit{ab initio} study by Palmino et al. \cite{REM1} and these
were very kindly provided by F. Palmino.

\begin{figure}[h]
\centering
  \includegraphics[width=3.5 in]{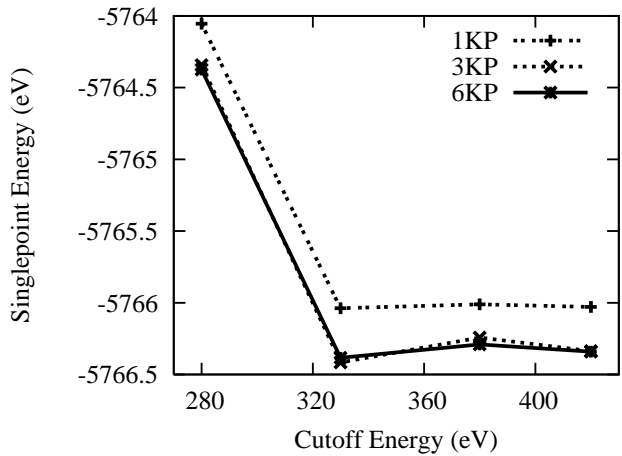}\\
  \caption{Variation of the singlepoint energy, which is the calculated energy
  for a given configuration of the atomic positions, with the cutoff energy and
  with the number of k-points at which the wavefunction is sampled in reciprocal
  space.}\label{fig:fig03}
\end{figure}

Before proceeding the input parameters in the calculation were
carefully checked (see \cite{Probert_Validation} for a discussion
of the importance of this). Figure \ref{fig:fig03} shows how the
calculated energy varies with the number of plane waves included
in the calculation as the cutoff energy is raised for three
increasingly dense Monkhorst-Pack \cite{KP} reciprocal space
sampling grids.

A cutoff energy of 380 eV yields a total energy that is
unambiguously in the variational minimum and will allow accurate
calculation of the energy and the forces within the system. We
have used the sampling grid with 3 k-points in reciprocal space
since an increase to 6 k-points does not significantly change the
energy. The Perdew-Burke-Ernzerhof \cite{PBE} generalised gradient
approximation was used to represent exchange and correlation
effects.

The vacuum gap that was used to prevent interaction between the
top surface in one supercell and the bottom surface in the
supercell above was 9 \AA \,thick and this has been optimised
during the course of other \textit{ab initio} studies of RE
silicides that we have done. We have included two bulk-like
silicon layers below the top layer that contains the samarium atom
and the honeycomb chain structure. To prevent interactions through
the supercell between uncompensated charge in the the top and
bottom layers and to fully replicate the transition to the bulk
silicon crystal we have hydrogen passivated the deepest bulk-like
silicon layer and fixed the coordinates of these atoms so that
they are not free to move from their bulk positions. We have
repeated the geometry optimisation of the unit cells without
passivation and positional constraints and the final positions of
the silicon atoms in this bottom layer are not drastically altered
and the total energy does not significantly change as a result
which suggests that using so few bulk-like layers is reasonable.
We nevertheless kept the hydrogen passivation in place since it
reduces the computational cost of the electronic structure
calculation by the quenching of dangling bonds on the underside of
the supercell.

\begin{figure}[t]
\centering \subfigure {
    \label{fig:fig04:a}
    \includegraphics[width=3.0 in]{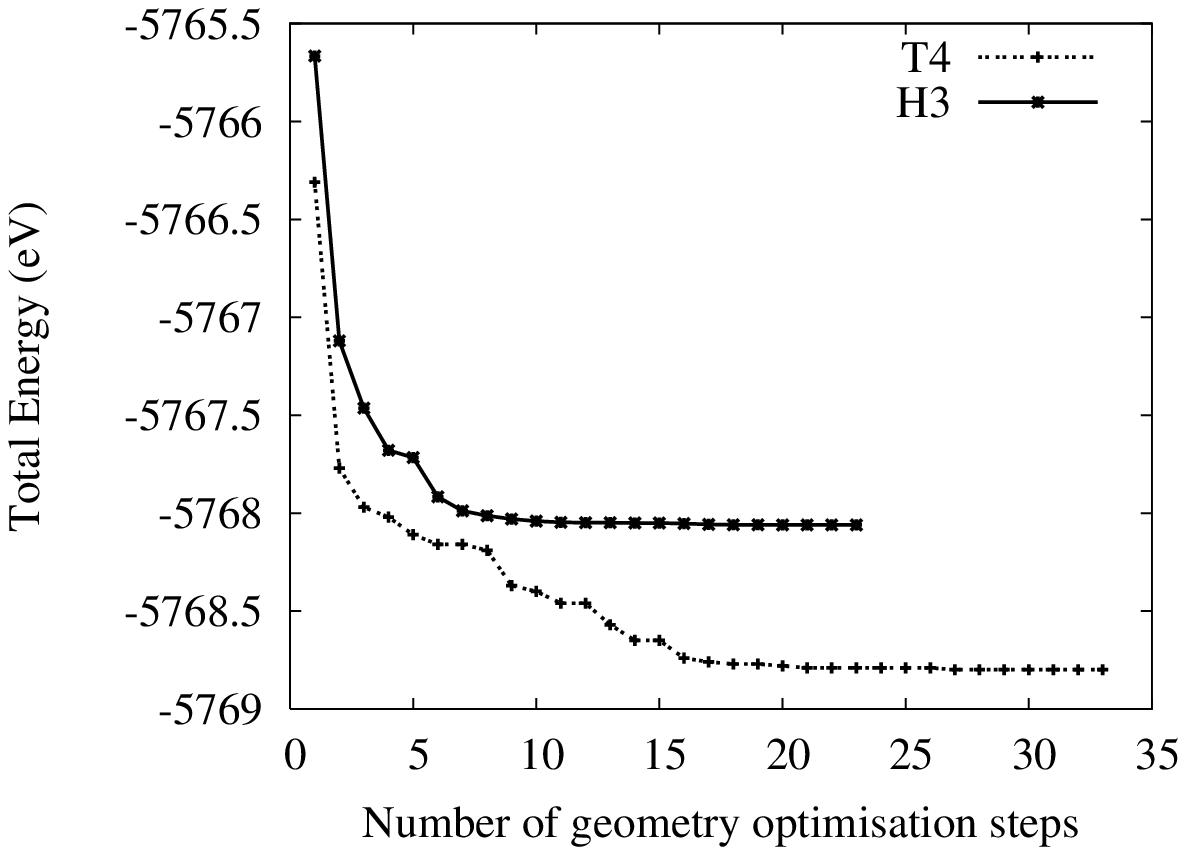}
} \hspace{1cm} \subfigure {
    \label{fig:fig04:b}
    \includegraphics[width=3.0 in]{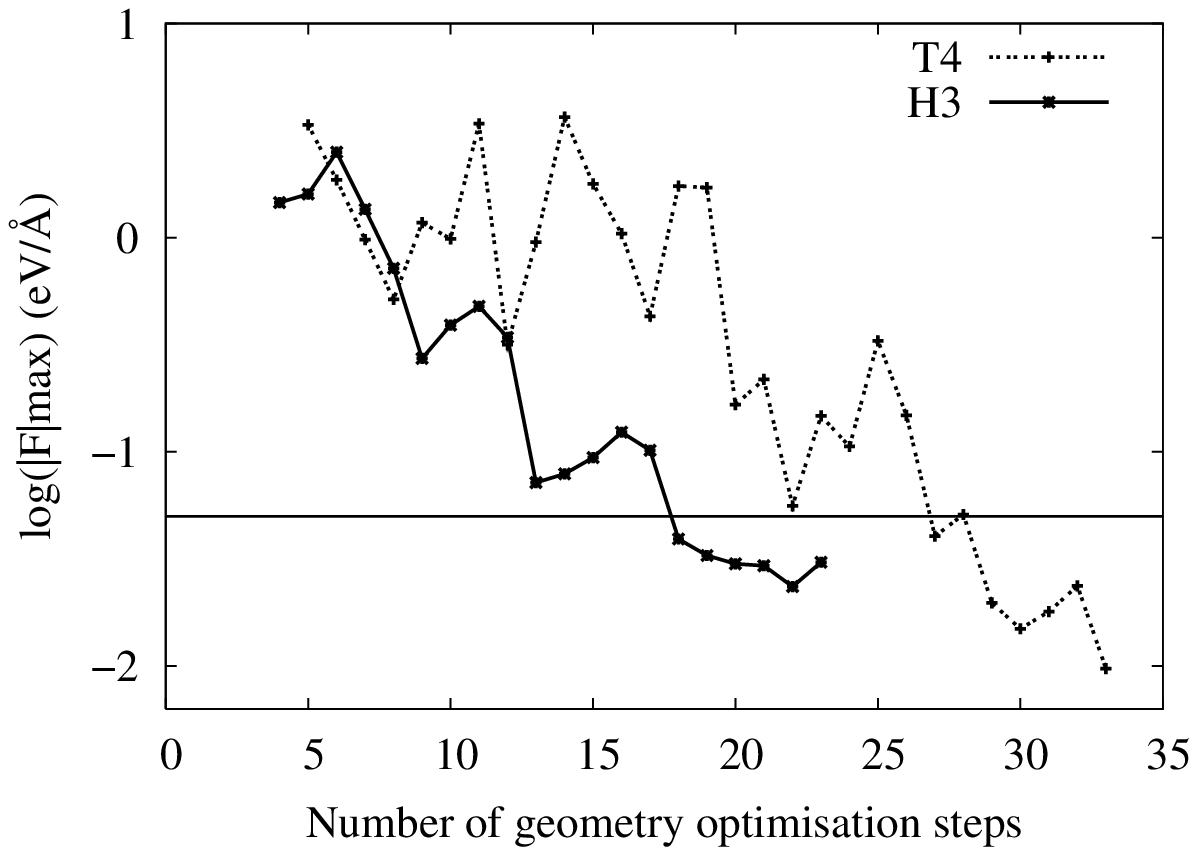}
}  \caption{Convergence of the total energy (top) and logarithmic
convergence of the forces (bottom) during the geometry
optimisation of the T4 and H3 structures. The horizontal line
indicates the force convergence tolerance of $5\times10^{-2}$
eV/\AA. The T4 structure has a lower energy than the H3 structure
and the maximum force on any atom is lower.}\label{fig:fig04}
\end{figure}

\begin{figure}[h]
\centering
  \includegraphics[width=3.5in]{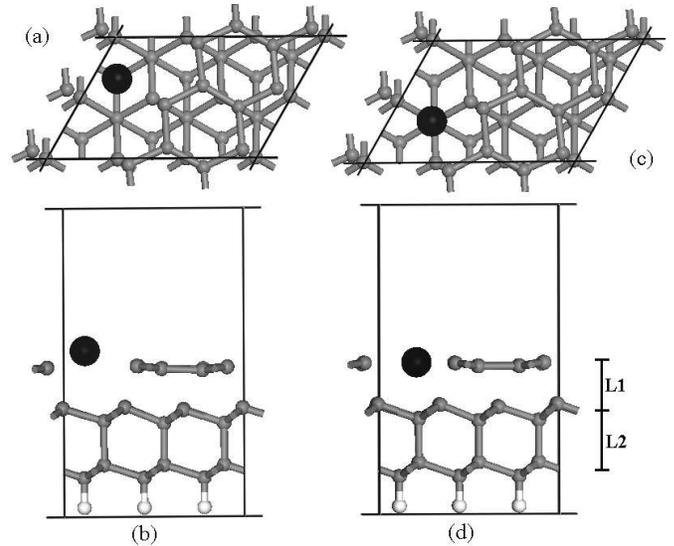}\\
  \caption{Optimised structures for the HCC model showing the H3
  model from above (a) and in side view (b) and the T4 model from
  above (c) and in side view (d). Silicon atoms here are grey, the
  samarium atom is black and the hydrogen atoms are white. The first
  and second interlayer spacings are labelled L1 and L2 respectively.
  }\label{fig:fig05}
\end{figure}

The structures were allowed to relax until the forces were below
the predefined tolerance of $5\times10^{-2}$ eV/\AA. Figure
\ref{fig:fig04} shows the convergence of the total energy and the
maximum force on any atom as the geometry optimisation proceeds
for the two structures.

The T4 structure is 0.7 eV (0.01\%) lower in energy than the H3
structure and the maximum force in the system is slightly lower.
This energy difference cannot be quantitatively compared with the
value of 0.07 eV/Sm that was obtained in Ref. \cite{REM1} since
this is an atomically resolved energy difference whereas the value
presented here compares the total energies of the two supercells
with contributions from all of the atoms within. Also, one cannot
compare the basis set parameters used in this work with those
presented in Ref. \cite{REM1} since the two calculations used
different types of pseudopotentials and a different \textit{ab
initio} code.

The final optimised structures are shown in figure
\ref{fig:fig05}. The interlayer spacings (ignoring the samarium
atom for now) in both structures here are almost identical. The
major difference between this calculated structure and that in
Ref. \cite{REM1} is in the interlayer spacings. In this study the
spacing between the top layer and the first bulk-like layer (L1 in
figure \ref{fig:fig05}) is approximately 8\% greater than that in
Ref. \cite{REM1} and the spacing between the first bulk-like layer
the the second bulk-like layer (L2 in figure \ref{fig:fig05}) is
about 4\% greater. There are also some minor differences in the
position of the silicon atoms in the honeycomb chain.

\section{IV. Comparison of Experiment and Theory}

\begin{figure}
\centering
  \includegraphics[width=3.0in]{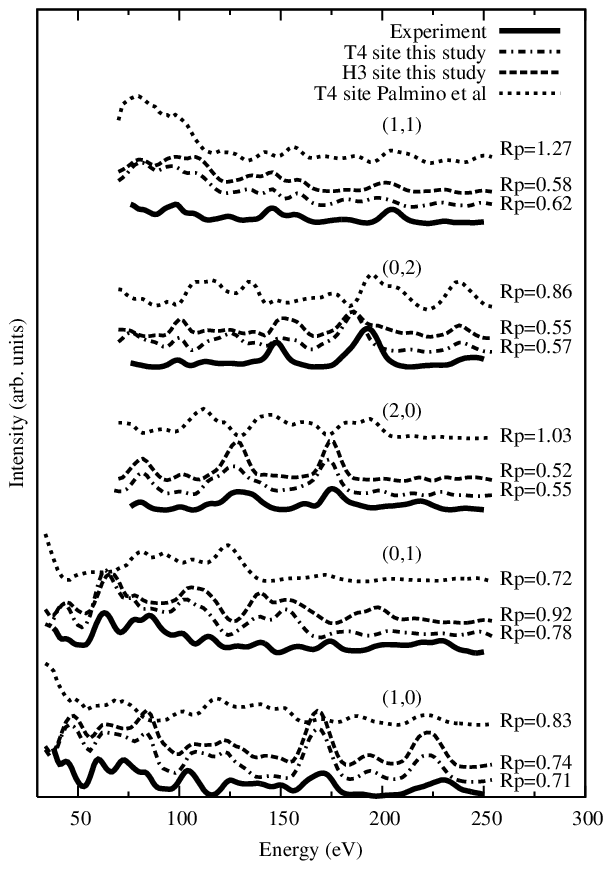}\\
  \caption{A comparison of the I-V curves calculated for the
  integer spots for the structures suggested by the
  \textit{ab initio} calculations in this study and elsewhere with
  those obtained experimentally. The R-factor beside each curve indicates
  the level of agreement with experiment.}\label{fig:fig06}
\end{figure}

Figure \ref{fig:fig06} shows I-V curves calculated using the
\texttt{CAVLEED} code \cite{CAVLEED} for the three candidate
\textit{ab initio} structures.  The curves shown are only the
integer spots in the LEED pattern and they were calculated using
the bulk Debye temperatures (that is 645 K for silicon and 169 K
for samarium) to represent the lattice vibrations of each layer.
The structures obtained from the two \textit{ab initio}
calculations in this study are a consistently better match to
experiment than that in Ref. \cite{REM1}. This suggests that the
interlayer spacings obtained in this study, to which LEED is very
sensitive, are closer to those present in the real surface. Also,
note that the I-V curves of the T4 and H3 structures from this
study are very similar and we cannot discard either structure.

We can divide the spots in the LEED pattern into two groups. The
integer spots ((1,0), (2,0), (1,1) etc) contain a large
contribution from the bulk and are sensitive to the top few
layers. The fractional spots ((2/3,0), (1/3,1) etc) are extremely
sensitive to the top layer reconstruction and only mildy sensitive
to deeper layers through multiple scattering.

The poor Pendry R-factors (that is $>$0.7 in this context where
enhanced vibrations have not been applied) for some of the integer
spots in figure \ref{fig:fig06} indicate that further structural
optimisation is needed. It is apparent that for some curves the
right peaks are present but that their energy is slightly wrong
(see the (0,2) and (2,0) spots in figure \ref{fig:fig06} for
example). The fractional spots have much better R-factors (see
figure~\ref{fig:fig07}) which indicates that the structure of the
top layer is in good agreement with experiment. The natural way to
proceed is to vary the interlayer spacings to attempt an
improvement in the match with experiment, particularly for the
integer spots. In the next section this is attempted.

\section{V. LEED I-V Structural Optimisation}

\begin{figure}
\flushright
    \subfigure{\label{fig:fig07:a} \includegraphics[width=3.5in]{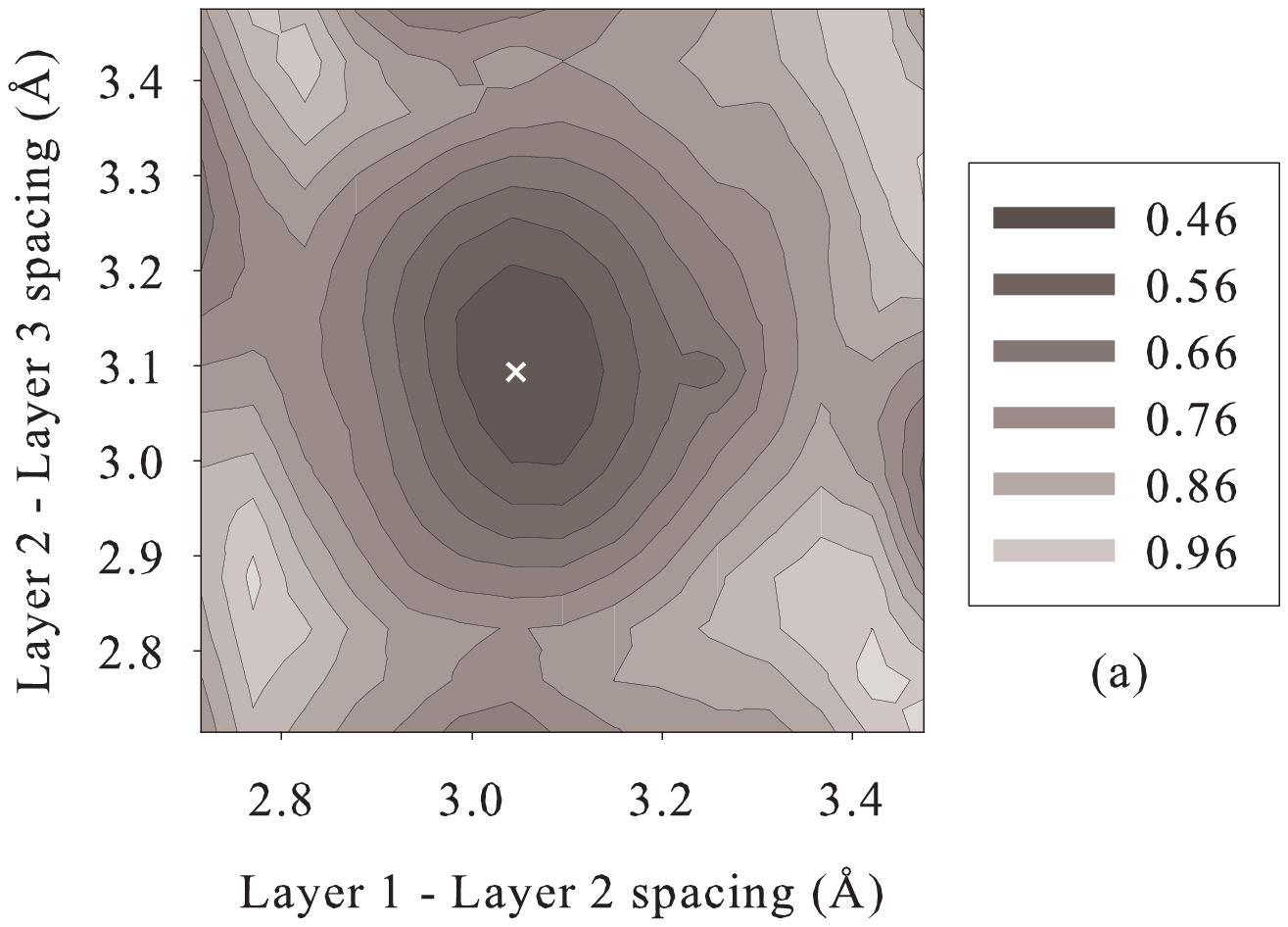}}
    \hspace{1cm}
    \subfigure{\label{fig:fig07:b} \includegraphics[width=3.5in]{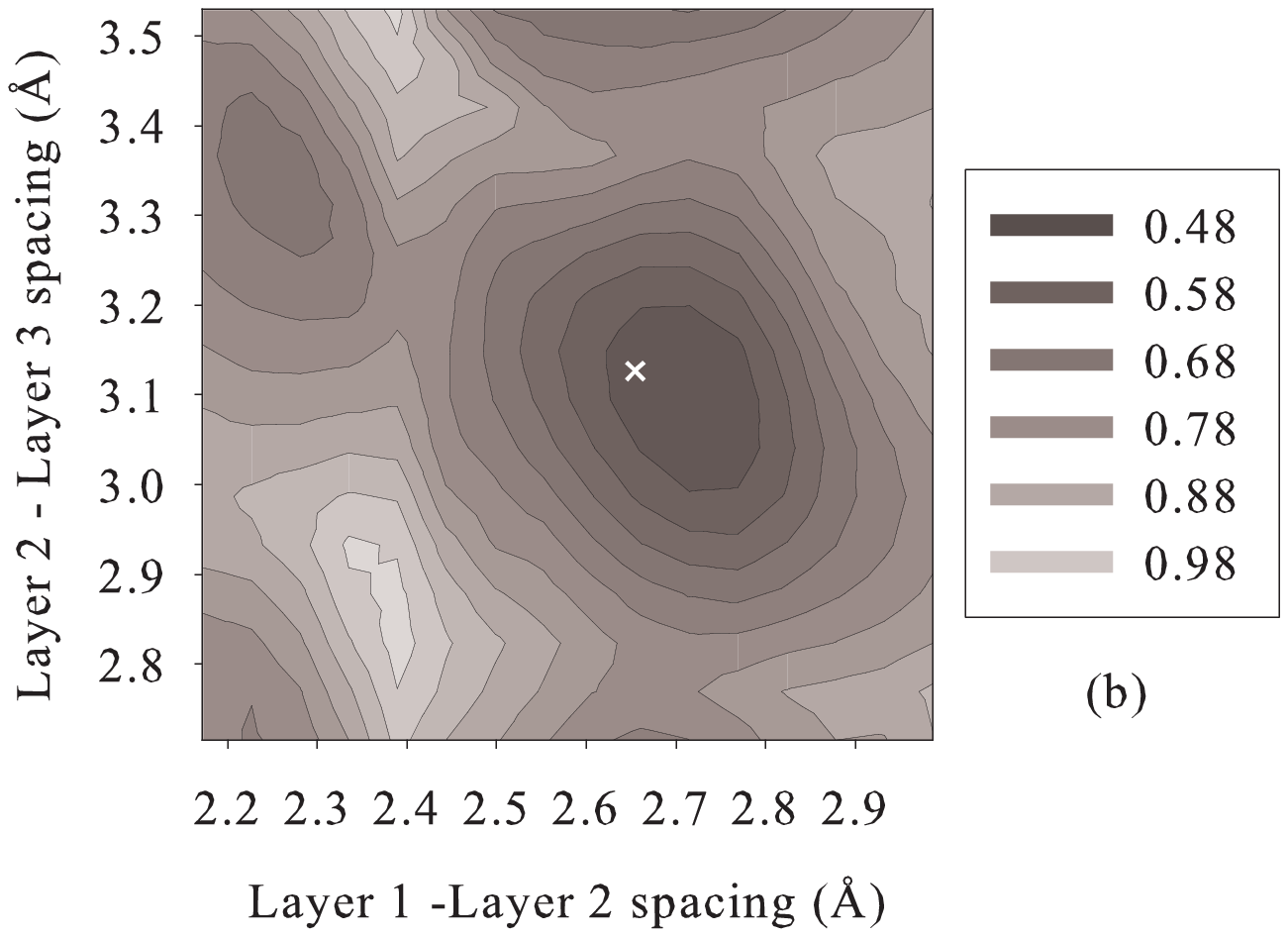}}
    \caption{Pendry R-factor landscape for a range of values of the
interlayer spacings in the (a) H3 and (b) T4 structure for the
fractional spots. The step size was 0.05 \AA. The cross indicates
the \textit{ab initio} energy minimum.} \label{fig:fig07}
\end{figure}

\noindent The calculation of the I-V curves was repeated using
various values for the interlayer spacings and the R-factors were
determined. An initial coarse search was carried out over a wide
range of values for the spacings and with a large step size.
Figure \ref{fig:fig07} shows the R-factor landscape obtained in
this manner for the fractional spots. There is a clear minimum in
both cases. The samarium atom has been considered in determining
the midpoint of the top layer which is why the minima do not
coincide; the samarium atom sits proud of the honeycomb layer in
the H3 structure and it is much lower in the T4 structure. In the
\textit{ab initio} calculations in this study the interlayer
spacings were approximately 3.06, 3.10\AA \,for the H3 structure
and 2.65, 3.14 \AA\, for the T4 structure which places the
\textit{ab initio} energy minimum (indicated by a cross in figure
\ref{fig:fig07}) very close to that of the \texttt{CAVLEED} I-V
R-factor minimum. Two independent techniques are thus suggesting
very similar best fit structures.

I-V curves were then obtained using a narrower range of interlayer
spacings focussed on the minima obtained in the coarse search.
This fine search, using a step size of 0.01 \AA ,\,improved the
R-factors by only around 0.01 in both cases and even finer
searches were not carried out.

\begin{figure}
\centering \subfigure{
    \label{fig:fig08:a}
    \includegraphics[width=2.4in]{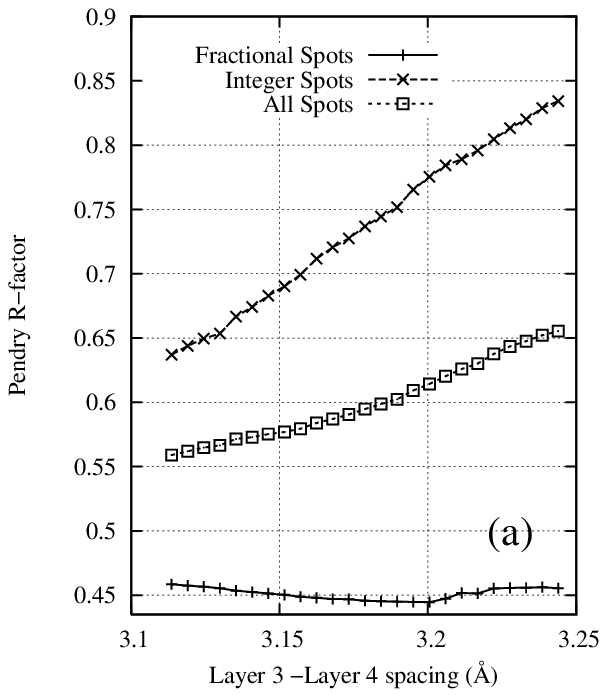}
} \hspace{1cm}
\subfigure 
{
    \label{fig:fig08:b}
    \includegraphics[width=2.4in]{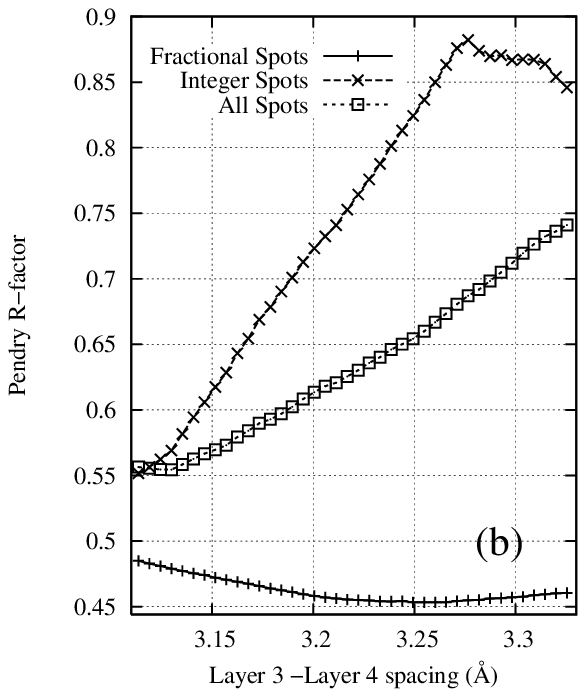}
} \caption{Variation of the spacing between layers three and four
  in Si(111)-3x2-Sm for the H3 (a) and T4 (b) structures. The
  bulk value for this interlayer spacing is 3.14\,\AA}
\label{fig:fig08}
\end{figure}

There is another interlayer spacing deeper into the bulk that we
might try to vary. Computational resources do not permit us to
independently vary this spacing along with those between the top
three layers. Figure \ref{fig:fig08} shows the variation of the
Pendry R-factor as the spacing between layers three and four is
changed with the first and second interlayer spacings fixed at
their optimum value. We can see that there is a small improvement
in the R-factor for the fractional spots at the expense of a large
worsening of the R-factor for the integer spots, which are more
sensitive to structure in the near bulk. We therefore reject any
variation of this interlayer spacing and retain the bulk value.
That there is no significant reconstruction deeper into the
surface justifies the use of three layers in our \textit{ab
initio} calculation and means that in both the \textit{ab initio}
calculation and the Pendry R-factor structure fit to the
experimental data we have considered two interlayer spacings.

\subsection{Optimisation of the vibrations used in the LEED I-V calculation}

\noindent The effect of thermal vibrations within the system has
also been investigated. The Debye temperature $T_{D}$ of the
samarium atom, the silicon atoms in the honeycomb layer and the
silicon atoms in the first bulk-like layer have each been
independently reduced by a factor of $\sqrt{2}, 2$ and $3$ from
their bulk values. The effects of these enhanced vibrations for
the two most effective combinations are shown in table
\ref{tab:table1} alongside the R-factors obtained with no enhanced
vibrations.

\begin{table}[h]

\begin{ruledtabular}
\begin{tabular}{ccccccc}
Sm $T_{D}$ & Si1 $T_{D}$ & Si2 $T_{D}$ & $R_{P}^{FRAC}$ & $R_{P}^{INT}$ & $R_{P}^{ALL}$\\
\hline
                                         &  &  & \\
  $B$                    & $B$   & $B$   & 0.49 & 0.72 & 0.63 \\
  $B/\sqrt{2}$           & $B/3$ & $B/2$ & 0.48 & 0.46 & 0.48 \\
  $B/2$                  & $B/3$ & $B/2$ & 0.45 & 0.44 & 0.45 \\
                                         & & & & & \\
\end{tabular}
\end{ruledtabular}

\caption{\label{tab:table1}Variation of the Debye temperature for
the samarium atom, silicon honeycomb layer and first bulk like
layer and the effect upon the Pendry R-factors for the H3
structure. The naming scheme here is Sm=samarium atom, Si1=silicon
honeycomb atoms, Si2=first silicon bulk-like layer. A Debye
temperature of $B$ indicates the bulk unoptimised value for that
atomic species. Similar data are available for the T4 structure.
Further enhancement of the vibrations of the samarium atom worsens
the R-factors.}
\end{table}

\noindent The two schemes of enhanced vibrations both reduce the
overall R-factor by around 0.2 and this is mainly due to the
improvement in the R-factors of the integer spots.

\subsection{Linear combination of the two candidate structures}

\begin{figure}[h]
\centering
\subfigure 
{
    \label{fig:fig09:a}
    \includegraphics[width=3.0 in]{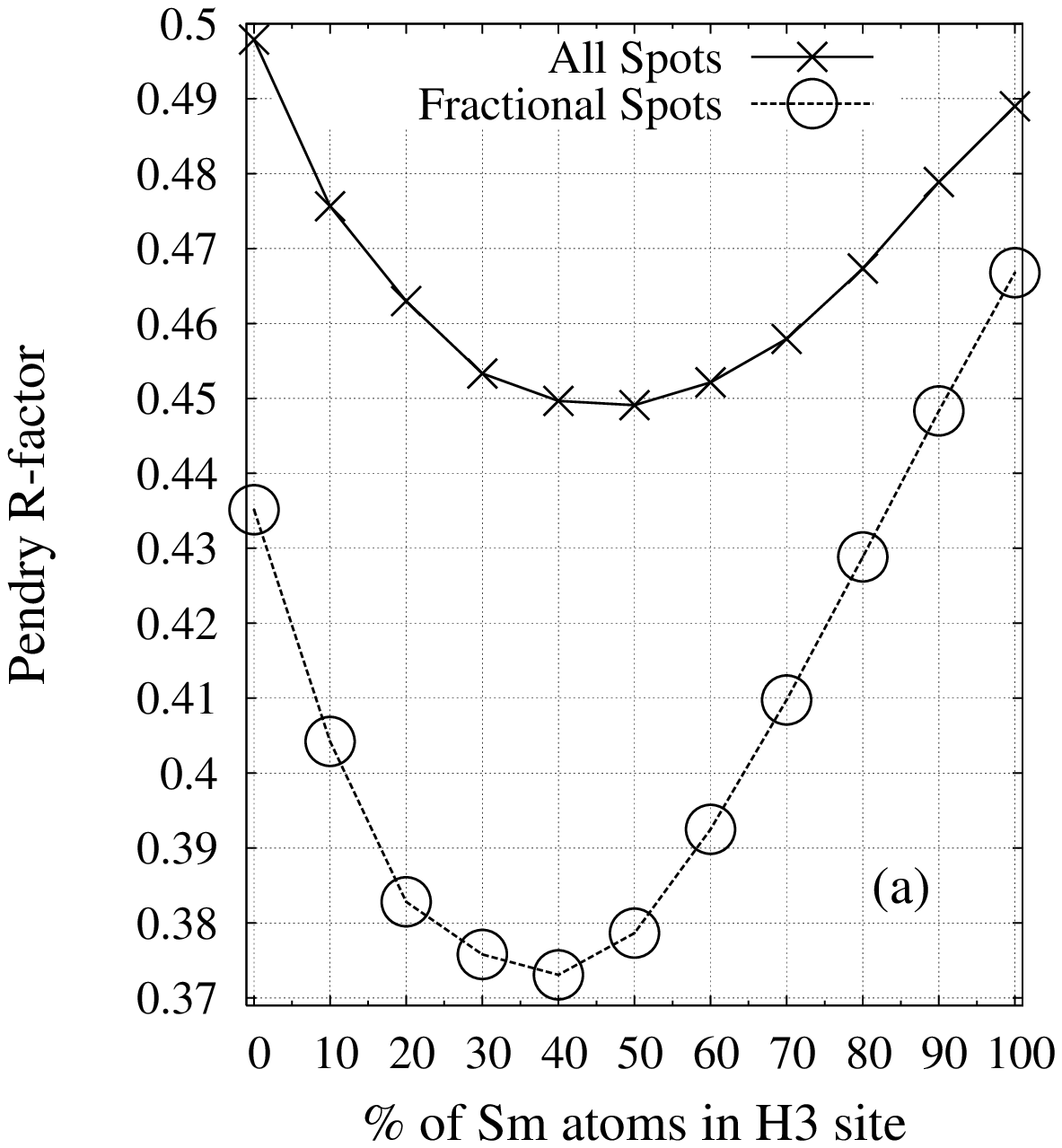}
} \hspace{1cm}
\subfigure 
{
    \label{fig:fig09:b}
    \includegraphics[width=3.0 in]{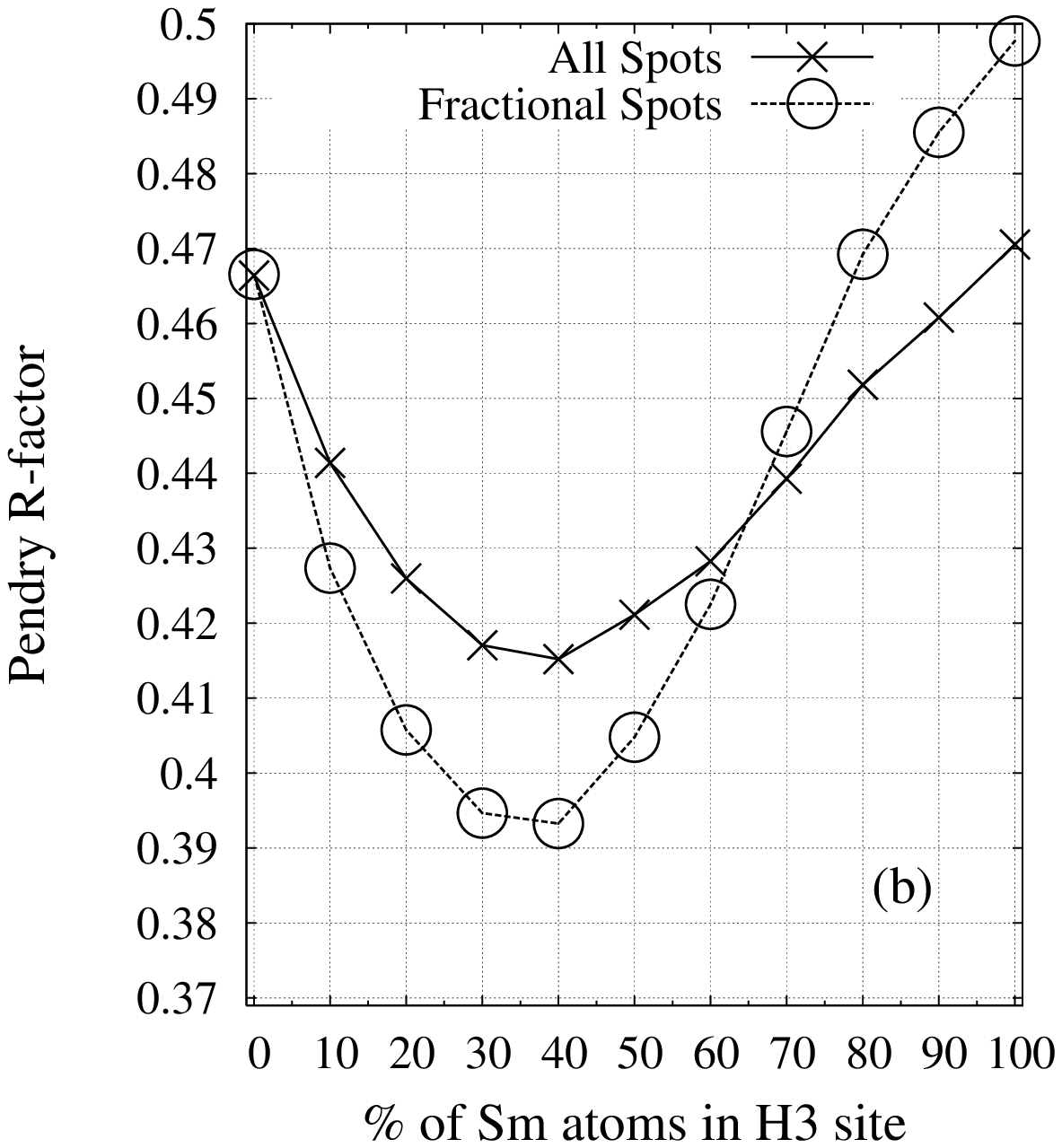}
} \caption{Pendry R-factors for a linear combination of the spot
intensities of the H3 and T4 structures in various mixing ratios
for two different vibrational regimes. In figure (a) the Debye
temperature for the samarium atom is 119 K  and in figure (b) it
is 84 K. In both cases the Debye temperatures of the top silicon
honeycomb layer, the first silicon bulk-like layer and the
repeated bulk layer are 215 K, 323 K and 645 K respectively.}
\label{fig:fig09}
\end{figure}

\noindent The H3 and T4 structures have similar energies, similar
structures (ignoring the position of the samarium atom) and
similar LEED I-V curves. It is reasonable to suggest that both
structures might co-exist upon the surface. A linear combination
of the I-V curves produced by the H3 and T4 structures that
individually best fit the experimental data are shown in figure
\ref{fig:fig09} for the two regimes of enhanced vibration shown in
table \ref{tab:table1}. The H3 and T4 structures are considered as
being separated by a distance greater than the coherence length of
the LEED beam. To simulate large and separate domains of the two
structures in this way the LEED spot intensities have been
combined and not the amplitudes.

The vibrational regime with a Debye temperature for the samarium
atom of 119 K ($B/\sqrt{2}$ in table \ref{tab:table1}) gives a
lower R-factor for the fractional order spots but it gives a worse
overall R-factor. The vibrational regime with a Debye temperature
for the samarium atom of 84 K ($B/2$ in table \ref{tab:table1})
gives a better overall R-factor and the minima for both the
fractional and the integer spots coincide. The final ratio of H3
40:60 T4 is in favour of the structure that is lower in energy
which is what we would expect.

\begin{figure}[h]
\centering
  \includegraphics[width=3.0in]{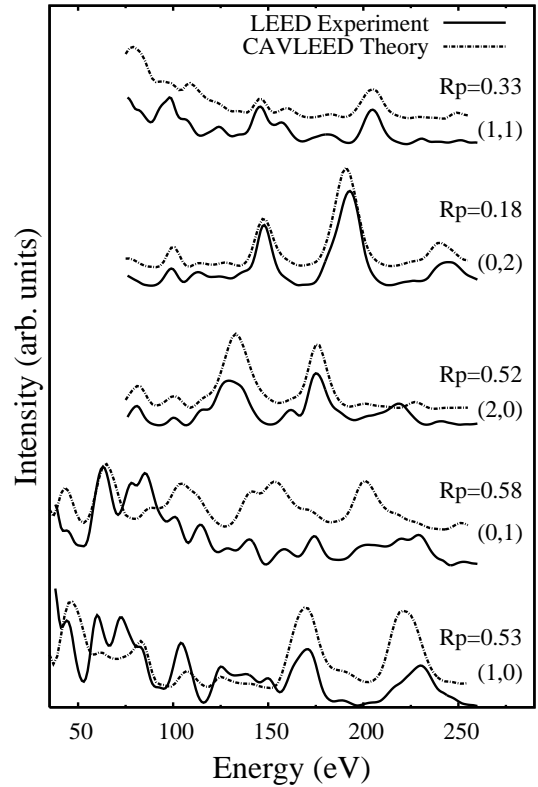}\\
  \caption{Best fit I-V curves for the integer LEED spots of the Si(111)-3x2-Sm structure.}\label{fig:fig10}
\end{figure}

\begin{figure}[h]
\centering
  \includegraphics[width=3.0in]{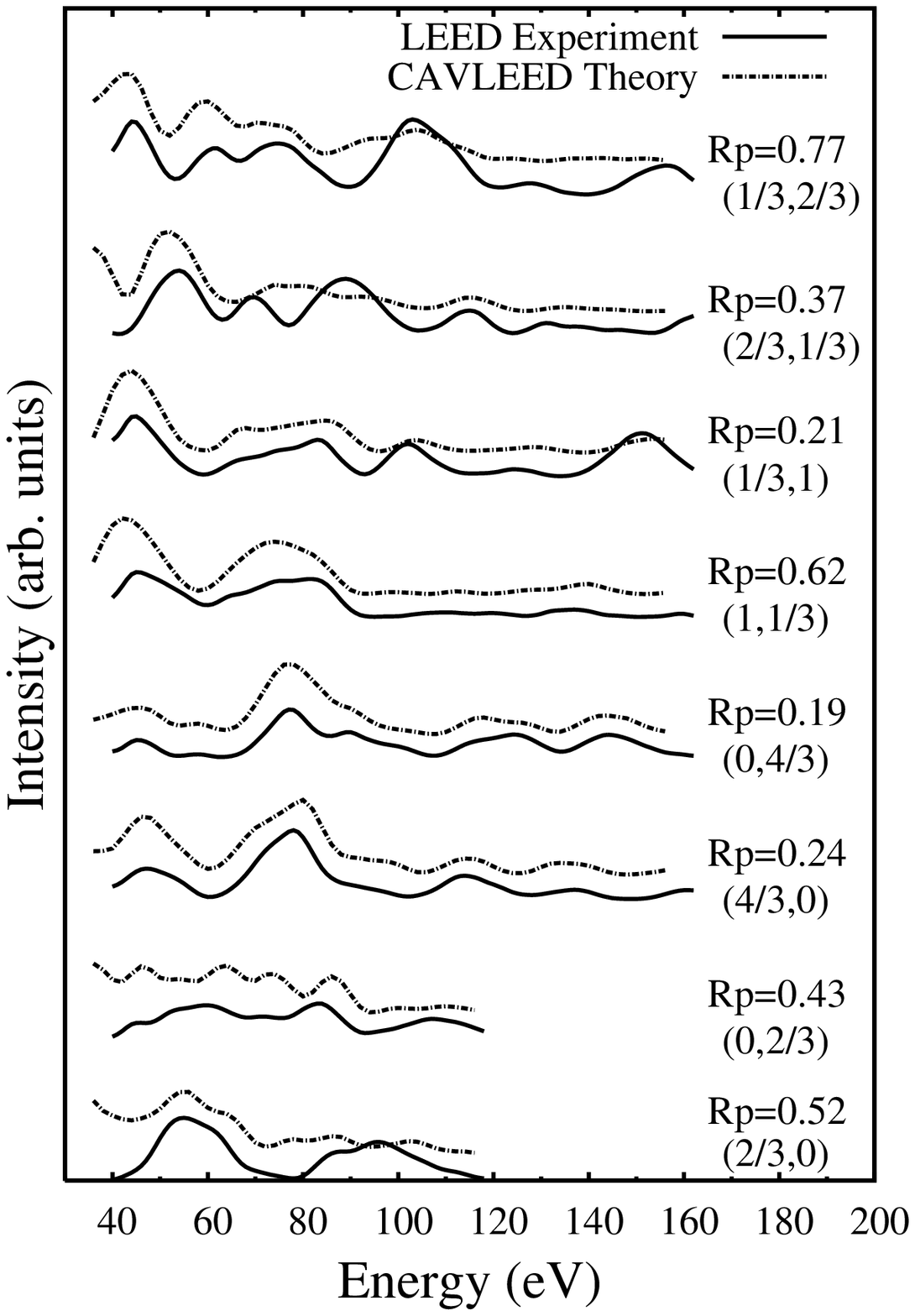}\\
  \caption{Best fit I-V curves for the fractional LEED spots of the Si(111)-3x2-Sm structure.}\label{fig:fig11}
\end{figure}

Table \ref{tab:table2} contains a summary of the structures
obtained from the \textit{ab initio} calculations and from the
\texttt{CAVLEED} LEED I-V structure fit. Two values are given for
L1; the value in brackets ignores the Sm atom in determining the
midpoint of the top layer. For the T4 structure the Sm atom is
almost coplanar with the honeycomb layer whereas for the H3
structure the Sm atom sits proud of the surface and skews the
value of L1. The value in brackets thus indicates the similarity
of the spacings between the layers of silicon atoms in the two
supercells. For each of the structures in this table LEED I-V
curves were calculated with optimised vibrations using a Debye
temperature for the samarium atom of 84 K. These were then
compared against experiment and the Pendry R-factors are included
in table \ref{tab:table2}. The final optimised LEED I-V curves for
the linear combination are compared with experiment for the
integer spots in figure \ref{fig:fig10} and for the fractional
spots in figure \ref{fig:fig11}.

\begin{table}[h]

\begin{ruledtabular}
\begin{tabular}{cccccc}
Structure & $R_{P}^{FRAC}$ & $R_{P}^{INT}$ & $R_{P}^{ALL}$ & L1 (\AA) & L2 (\AA)\\
\hline
                               &  &  & \\
  T4 (Ref. \cite{REM1})        & 0.87 & 0.88 & 0.92 & 2.42 (2.52) & 3.02 \\
  T4 \texttt{CASTEP}           & 0.44 & 0.46 & 0.46 & 2.65 (2.67) & 3.14 \\
  H3 \texttt{CASTEP}           & 0.47 & 0.43 & 0.45 & 3.06 (2.62) & 3.10 \\
  T4 \texttt{CAVLEED}          & 0.48 & 0.41 & 0.46 & 2.74 (2.73) & 3.10 \\
  H3 \texttt{CAVLEED}          & 0.45 & 0.44 & 0.45 & 3.06 (2.64) & 3.11 \\
  Combination           & \textbf{0.39} & \textbf{0.42} & \textbf{0.41} & \textbf{2.87 (2.69)} & \textbf{3.10} \\
                               & & & & & \\
\end{tabular}
\end{ruledtabular}

\caption{\label{tab:table2} Pendry R-factors for the fractional
spots ($R_{P}^{FRAC}$), integer spots ($R_{P}^{INT}$) and for all
spots ($R_{P}^{ALL}$) for the various optimised structures in this
work. All of the calculated I-V curves used optimised vibrations.
The interlayer spacings are shown in columns five and six
midpoint. The value of L1 in brackets ignores the Sm atom in the
determination of the midpoint of the top layer and indicates the
similarity of the structure of the silicon atoms in the two
supercells.}
\end{table}

\section{VI. LEED I-V Investigation of the missing half order spots}

\begin{figure}
\centering \subfigure[\,\,\,Integer spots] {
    \label{fig:fig12:a}
    \includegraphics[width=2.5 in]{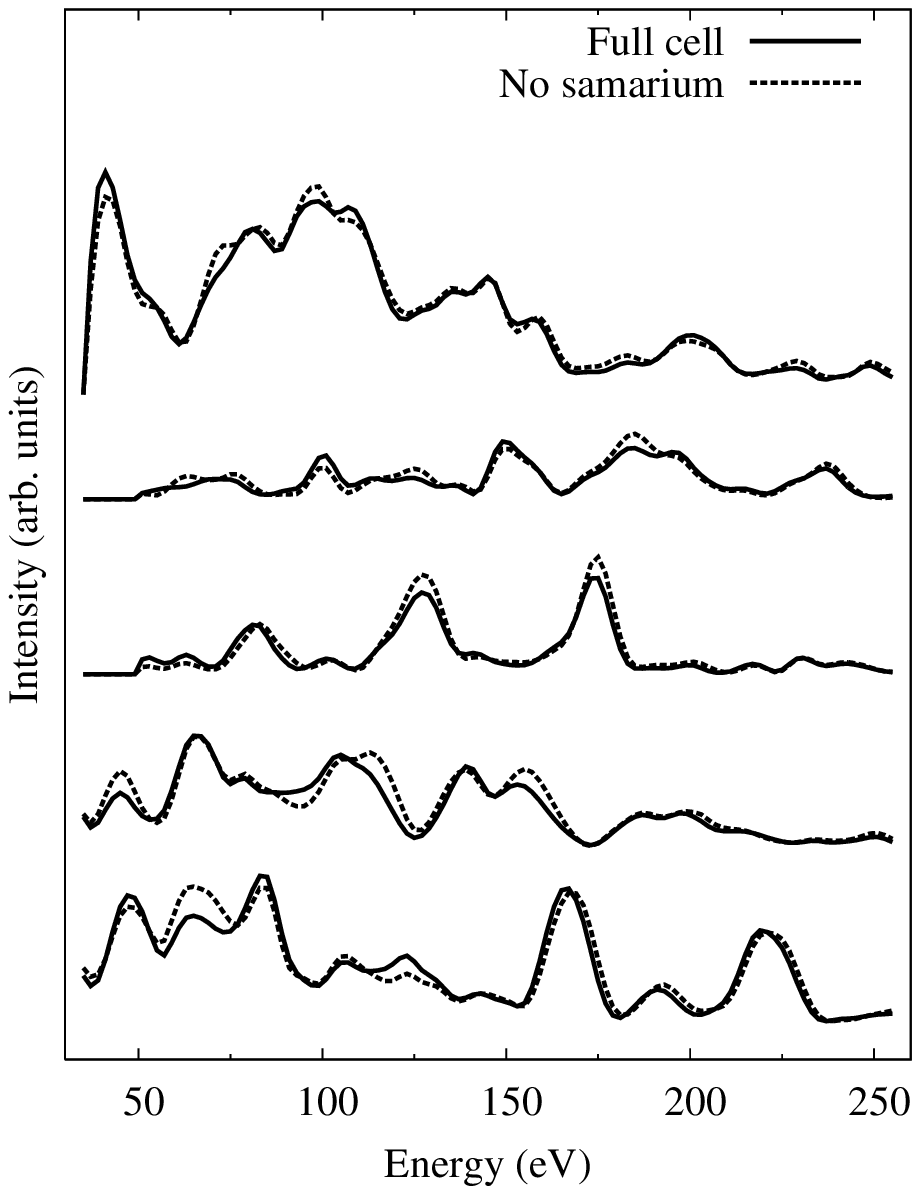} } \hspace{1cm}
\subfigure[\,\,\,Fractional spots] {
    \label{fig:fig12:b}
    \includegraphics[width=2.5 in]{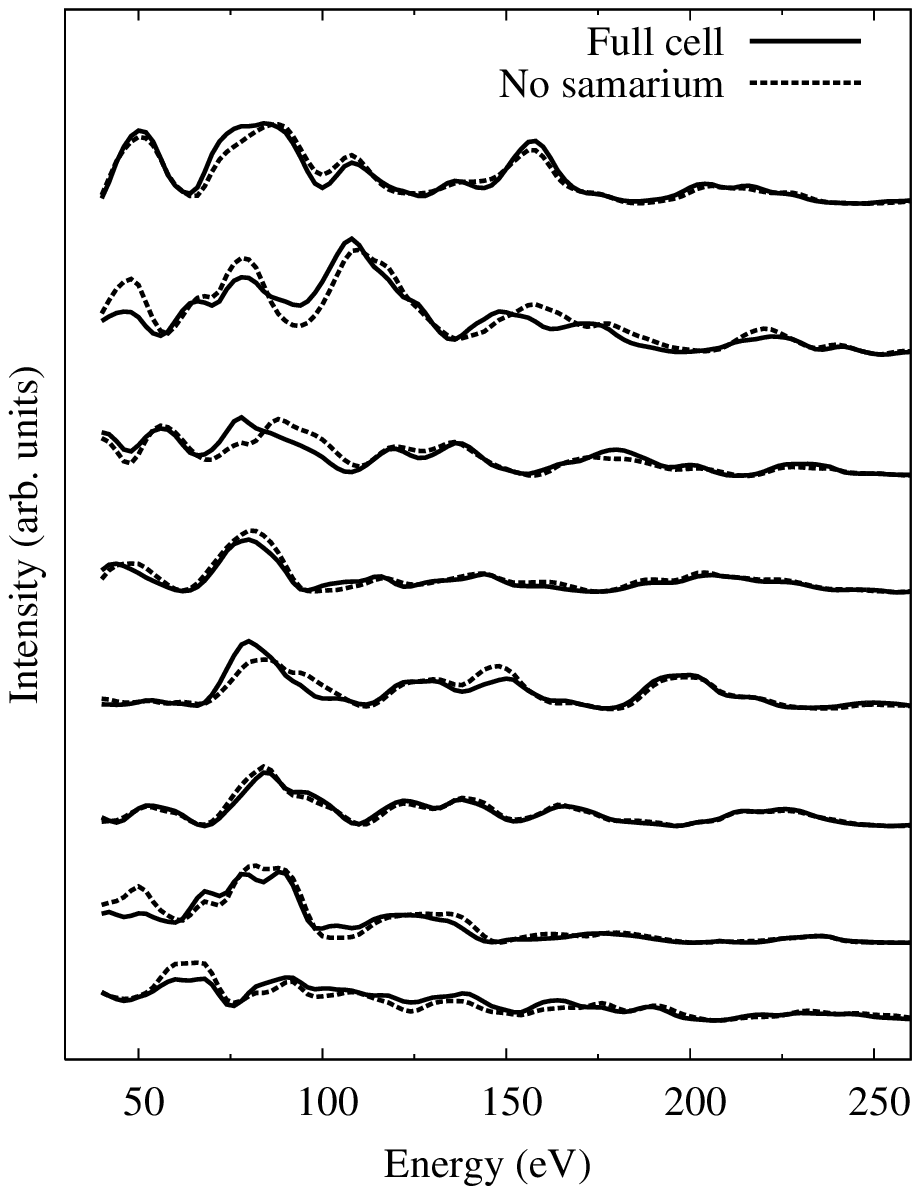}
} \caption{Calculated LEED I-V curves for the integer spots (a)
and fractional spots (b) of the H3 unit cell with and without the
samarium atom in place. Bulk Debye temperatures were used
throughout.} \label{fig:fig12}
\end{figure}

\noindent The silicon honeycomb layer is almost mirror symmetric
about a plane perpendicular to the $\times$2 direction. It is the
location of the samarium atom that breaks this mirror symmetry and
renders a quasi 3$\times$1 unit cell into a 3$\times$2 unit cell.
Figure \ref{fig:fig12} shows calculated I-V curves for the H3
structure for the fractional and integer spots compared with those
for the same structure with the samarium atom removed. The bulk
Debye temperatures were used throughout to minimise the influence
of vibrations. It is readily apparent that the I-V curves are
insensitive to the presence of the samarium atom.

This is not to say that the samarium atom is not a strong
scatterer. It would appear that the silicon honeycomb layer as a
scattering unit of 8 atoms contributes much more to the I-V curves
than the single samarium atom. A similar effect was observed in
the LEED I-V structural analysis of Ag- and Li-induced
Si(111)-($\sqrt{3} \times \sqrt{3} )R30^{\,\circ}$ by Over et al.
\cite{Fan2} and was suggested as a cause for the
3$\times$1/3$\times$2 discrepancy in Ref. \cite{3xn7}.

\begin{figure}[h]
\centering
  \includegraphics[width=3in]{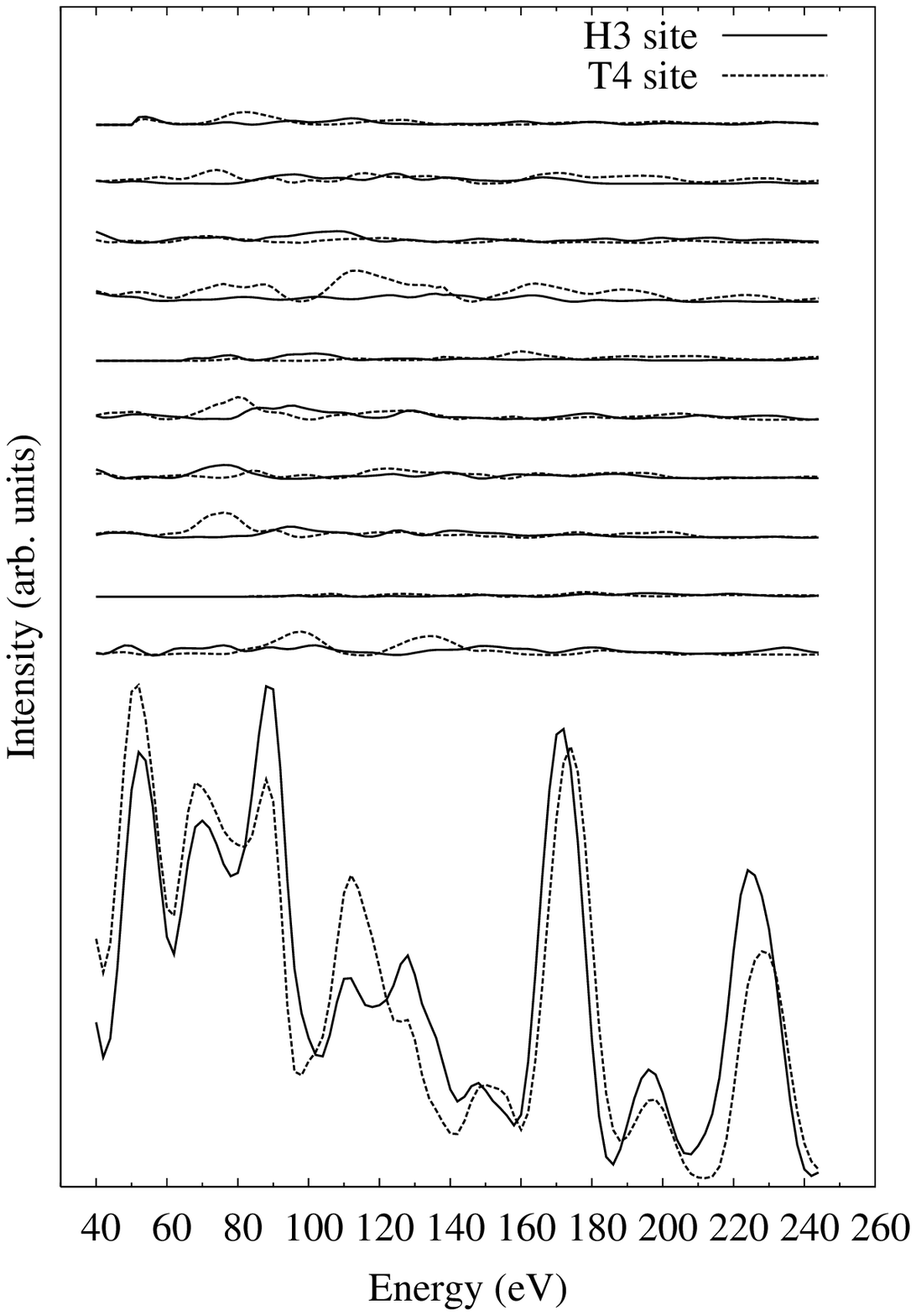}\\
  \caption{Calculated I-V curves for the HCC structure showing the difference in
  the intensity (typically an order of magnitude) between the
  half order spots and a representative spot that is visible in the LEED pattern
  during an experiment.}\label{fig:fig13}
\end{figure}

If this is the case then the half order spots that are apparently
missing when the experimental 3$\times$1 LEED pattern is inspected
visually should produce calculated I-V curves whose intensity is
very much less than that of the spots that are visible during
experiment. The silicon honeycomb layer is not perfectly
symmetrical about the mirror plane perpendicular to the $\times$2
direction and this should contribute to the half order spot
intensities. Figure \ref{fig:fig13} shows the I-V curves of some
of the calculated half order spots compared to that of the (1,0)
spot.

It would appear that the 3$\times$2 unit cell produces a
3$\times$2 LEED pattern with half order spots that are so weak in
intensity that they fall below the background intensity leaving
only a 3$\times$1 LEED pattern visible.

\section{VII. Discussion}

The Pendry R-factors obtained upon comparison of the \textit{ab
initio} calculations with experiment are not as low as we would
expect. We can see that for some spots the I-V curves are visually
very similar to those obtained experimentally (see the I-V curves
for the (1,1/3) and (2,0) spots for example) but they have a poor
R-factor. This suggests that the structure is very nearly right
and the minor discrepancy could be a result of our not including
enough bulk like silicon layers in the bottom of the supercell
with consequent effects upon the reconstruction within the top
honeycomb layer. We have attempted some simple variation in the
top layer structure, for example flattening the layer, but this
drastically worsens the R-factor. Computational resources prohibit
us from calculating the structure with more silicon layers and
from investigating the honeycomb layer structure further using
LEED I-V and perhaps further study with a LEED I-V genetic
algorithmn search might optimise this structure further. The
moderate R-factors are offset by the fact that two independent
techniques both show optimum structural fits for almost identical
interlayer spacings.

The lateral atomic structure was freely varied in the \textit{ab
initio} calculations in this work and the lateral atomic positions
agree well with those found by Palmino at al. \cite{REM1} which
they have shown to be in good qualitative agreement with
experimental STM images. In this work we have concentrated upon
the optimisation of the vertical spacings, to which LEED is
particularly sensitive.

The R-factors for the integer spots are consistently worse than
those for the fractional spots. There is the possibility that
there are some regions in which there is a disordered overlayer of
samarium atop a bulk terminated Si(111)-1$\times$1 surface. Such a
phase has been reported by Wigren et al. \cite{Wigren2}. The
integer spots from such regions might contribute to the overall
integer spots for the surface and reduce the level of agreement
with the calculated I-V curves for the pure 3$\times$2 surface.

We have not been able to determine the long range order in the
system. We might expect that simple electrostatic repulsion along
the 1D chain would space out the metal atoms and provide large
separate domains of the H3 and the T4 structures. However, the two
sites are almost degenerate and there would be an entropic gain
from disorder. In the literature one can find evidence for both
order and disorder in the long range positions of the metal atoms.
In this study the improvement in the Pendry R-factor when the T4
and H3 structures are considered together on the surface suggests
that both sites are occupied within the surface. We have also
shown that we do not require more than one unit cell to explain
the missing half order spots in the LEED pattern and our
experimentally observed LEED patterns show a low background due to
good order on the surface. It could be that there is long range
disorder on the surface and that the coupling between many
adjacent H3 and T4 unit cells and matching of the interlayer
spacings introduces a slight strain that changes the structure in
the honeycomb layer and the first bulk-like layer enough to
account for our Pendry R-factors. If this is the case then it
would be impossible to obtain the structure of the honeycomb layer
to a high degree of accuracy without an \textit{ab initio}
calculation using a supercell that comprises several thousand unit
cells of the H3 and T4 structures randomly tesselated in both
directions.

\section{VIII. Summary}

We have provided a quantitative validation of the honeycomb-chain
channel model common to the 3$\times$1 and 3$\times$2 structures
formed by alkali, alkali-earth and rare-earth metals on Si(111).
Several I-V datasets were obtained from LEED experiments and used
to fingerprint the surface. The atomic structure suggested by our
two \textit{ab initio} calculations is in reasonable agreement
with this experimental data. Further structural optimisation and
mapping of the R-factor landscape have shown that a slight outward
expansion of the top layer improves the fit somewhat but
increasing the vibrations in the top two layers gives a
significant improvement. A linear combination of the two HCC
structures has been shown to improve the fit still further with
the ratio being slightly in favour of the structure with the lower
energy of the two. Finally, we have calculated the intensities of
the half order spots and shown that they are sufficiently dim to
fall below the background intensity in a LEED experiment. Little
change in the calculated I-V curves results from removing the
samarium atom which supports the idea that as a scattering unit
the silicon honeycomb layer dominates the unit cell and makes LEED
insensitive to the metal atom in these 3$\times$ systems.

\section{IX. Acknowledgements}

\noindent Many thanks to F. Palmino for kindly providing the
atomic co-ordinates for the Si(111)-3$\times$2 unit cell.\\

C. Eames would like to acknowledge the EPSRC for financial
support.\\

This work made use of the facilities of HPCx, the UK's national
high-performance computing service, which is provided by EPCC at
the University of Edinburgh and by CCLRC Daresbury Laboratory, and
funded by the Office of Science and Technology through EPSRC's
High End Computing Programme.


\begin{thebibliography}{26}
\expandafter\ifx\csname
natexlab\endcsname\relax\def\natexlab#1{#1}\fi
\expandafter\ifx\csname bibnamefont\endcsname\relax
  \def\bibnamefont#1{#1}\fi
\expandafter\ifx\csname bibfnamefont\endcsname\relax
  \def\bibfnamefont#1{#1}\fi
\expandafter\ifx\csname citenamefont\endcsname\relax
  \def\citenamefont#1{#1}\fi
\expandafter\ifx\csname url\endcsname\relax
  \def\url#1{\texttt{#1}}\fi
\expandafter\ifx\csname
urlprefix\endcsname\relax\def\urlprefix{URL }\fi
\providecommand{\bibinfo}[2]{#2}
\providecommand{\eprint}[2][]{\url{#2}}

\bibitem[{\citenamefont{Saranin et~al.}(1998)\citenamefont{Saranin, Zotov,
  Ryzhkov, Tsukanov, Lifshits, Ryu, Kubo, Tani, Harada, Katayama and Oura.}}]{AM1}
\bibinfo{author}{\bibfnamefont{A.~A.} \bibnamefont{Saranin}},
  \bibinfo{author}{\bibfnamefont{A.~V.} \bibnamefont{Zotov}},
  \bibinfo{author}{\bibfnamefont{S.~V.} \bibnamefont{Ryzhkov}},
  \bibinfo{author}{\bibfnamefont{D.~A.} \bibnamefont{Tsukanov}},
  \bibinfo{author}{\bibfnamefont{V.~G.} \bibnamefont{Lifshits}},
  \bibinfo{author}{\bibfnamefont{J.-T.} \bibnamefont{Ryu}},
  \bibinfo{author}{\bibfnamefont{O.}~\bibnamefont{Kubo}},
  \bibinfo{author}{\bibfnamefont{H.}~\bibnamefont{Tani}},
  \bibinfo{author}{\bibfnamefont{T.}~\bibnamefont{Harada}},
  \bibinfo{author}{\bibfnamefont{M.}~\bibnamefont{Katayama}}, \bibnamefont{and}
  \bibinfo{author}{\bibfnamefont{K.}~\bibnamefont{Oura}}, \bibinfo{journal}{Phys. Rev. B.,}
  \textbf{\bibinfo{volume}{58}}, \bibinfo{pages}{7059} (\bibinfo{year}{1998}).

\bibitem[{\citenamefont{Weitering et~al.}(1996)\citenamefont{Weitering, Shi,
  and Erwin}}]{AM2}
\bibinfo{author}{\bibfnamefont{H.~H.} \bibnamefont{Weitering}},
  \bibinfo{author}{\bibfnamefont{X.}~\bibnamefont{Shi}}, \bibnamefont{and}
  \bibinfo{author}{\bibfnamefont{S.~C.} \bibnamefont{Erwin}},
  \bibinfo{journal}{Phys. Rev. B.,} \textbf{\bibinfo{volume}{54}},
  \bibinfo{pages}{10585} (\bibinfo{year}{1996}).

\bibitem[{\citenamefont{Sakamoto et~al.}(2002)\citenamefont{Sakamoto, Takeyama,
  Zhang, and Uhrberg}}]{AEM1}
\bibinfo{author}{\bibfnamefont{K.}~\bibnamefont{Sakamoto}},
  \bibinfo{author}{\bibfnamefont{W.}~\bibnamefont{Takeyama}},
  \bibinfo{author}{\bibfnamefont{H.~M.} \bibnamefont{Zhang}}, \bibnamefont{and}
  \bibinfo{author}{\bibfnamefont{R.~I.~G.} \bibnamefont{Uhrberg}},
  \bibinfo{journal}{Phys. Rev. B.,} \textbf{\bibinfo{volume}{66}},
  \bibinfo{pages}{165319} (\bibinfo{year}{2002}).

\bibitem[{\citenamefont{Lee et~al.}(2001)\citenamefont{Lee, Hong, Kim, Shin,
  Koo, Lee, and Moon}}]{AEM2}
\bibinfo{author}{\bibfnamefont{G.}~\bibnamefont{Lee}},
  \bibinfo{author}{\bibfnamefont{S.}~\bibnamefont{Hong}},
  \bibinfo{author}{\bibfnamefont{H.}~\bibnamefont{Kim}},
  \bibinfo{author}{\bibfnamefont{D.}~\bibnamefont{Shin}},
  \bibinfo{author}{\bibfnamefont{J.-Y.} \bibnamefont{Koo}},
  \bibinfo{author}{\bibfnamefont{H.-I.} \bibnamefont{Lee}}, \bibnamefont{and}
  \bibinfo{author}{\bibfnamefont{D.~W.} \bibnamefont{Moon}},
  \bibinfo{journal}{Phys. Rev. Lett.,} \textbf{\bibinfo{volume}{87}},
  \bibinfo{pages}{56104} (\bibinfo{year}{2001}).

\bibitem[{\citenamefont{Palmino et~al.}(2003)\citenamefont{Palmino, Ehret,
  Mansour, Labrune, Lee, Kim, and Themlin}}]{REM1}
\bibinfo{author}{\bibfnamefont{F.}~\bibnamefont{Palmino}},
  \bibinfo{author}{\bibfnamefont{E.}~\bibnamefont{Ehret}},
  \bibinfo{author}{\bibfnamefont{L.}~\bibnamefont{Mansour}},
  \bibinfo{author}{\bibfnamefont{J.-C.} \bibnamefont{Labrune}},
  \bibinfo{author}{\bibfnamefont{G.}~\bibnamefont{Lee}},
  \bibinfo{author}{\bibfnamefont{H.}~\bibnamefont{Kim}}, \bibnamefont{and}
  \bibinfo{author}{\bibfnamefont{J.-M.} \bibnamefont{Themlin}},
  \bibinfo{journal}{Phys. Rev. B.,} \textbf{\bibinfo{volume}{67}},
  \bibinfo{pages}{195413} (\bibinfo{year}{2003}).

\bibitem[{\citenamefont{Kuzmin et~al.}(2003)\citenamefont{Kuzmin, Vaara,
  Laukkanen, Per{\"a}l{\"a}, and V{\"a}yrynen}}]{REM2}
\bibinfo{author}{\bibfnamefont{M.}~\bibnamefont{Kuzmin}},
  \bibinfo{author}{\bibfnamefont{R.-L.} \bibnamefont{Vaara}},
  \bibinfo{author}{\bibfnamefont{P.}~\bibnamefont{Laukkanen}},
  \bibinfo{author}{\bibfnamefont{R.}~\bibnamefont{Per{\"a}l{\"a}}},
  \bibnamefont{and} \bibinfo{author}{\bibfnamefont{I.~J.}
  \bibnamefont{V{\"a}yrynen}}, \bibinfo{journal}{Surf. Sci.,}
  \textbf{\bibinfo{volume}{538}}, \bibinfo{pages}{124} (\bibinfo{year}{2003}).

\bibitem[{\citenamefont{Weitering}(1996)}]{3xn1}
\bibinfo{author}{\bibfnamefont{H.~H.} \bibnamefont{Weitering}},
  \bibinfo{journal}{Surf. Sci.,} \textbf{\bibinfo{volume}{355}},
  \bibinfo{pages}{L271} (\bibinfo{year}{1996}).

\bibitem[{\citenamefont{Jeon et~al.}(1992)\citenamefont{Jeon, Hashizume,
  Sakurai, and Willis}}]{3xn2}
\bibinfo{author}{\bibfnamefont{D.}~\bibnamefont{Jeon}},
  \bibinfo{author}{\bibfnamefont{T.}~\bibnamefont{Hashizume}},
  \bibinfo{author}{\bibfnamefont{T.}~\bibnamefont{Sakurai}}, \bibnamefont{and}
  \bibinfo{author}{\bibfnamefont{R.~F.} \bibnamefont{Willis}},
  \bibinfo{journal}{Phys. Rev. Lett.,} \textbf{\bibinfo{volume}{69}},
  \bibinfo{pages}{1419} (\bibinfo{year}{1992}).

\bibitem[{\citenamefont{Wigren et~al.}(1993{\natexlab{a}})\citenamefont{Wigren,
  Andersen, Nyholm, Gothelid, Hammar, Tornevik, and Karlson}}]{3xn3}
\bibinfo{author}{\bibfnamefont{C.}~\bibnamefont{Wigren}},
  \bibinfo{author}{\bibfnamefont{J.~N.} \bibnamefont{Andersen}},
  \bibinfo{author}{\bibfnamefont{R.}~\bibnamefont{Nyholm}},
  \bibinfo{author}{\bibfnamefont{M.}~\bibnamefont{Gothelid}},
  \bibinfo{author}{\bibfnamefont{M.}~\bibnamefont{Hammar}},
  \bibinfo{author}{\bibfnamefont{C.}~\bibnamefont{Tornevik}}, \bibnamefont{and}
  \bibinfo{author}{\bibfnamefont{U.~O.} \bibnamefont{Karlson}},
  \bibinfo{journal}{Phys. Rev. B.,} \textbf{\bibinfo{volume}{48}},
  \bibinfo{pages}{11014} (\bibinfo{year}{1993}{\natexlab{a}}).

\bibitem[{\citenamefont{Quinn and Jona}(1991)}]{3xn4}
\bibinfo{author}{\bibfnamefont{J.}~\bibnamefont{Quinn}} \bibnamefont{and}
  \bibinfo{author}{\bibfnamefont{F.}~\bibnamefont{Jona}},
  \bibinfo{journal}{Surf. Sci.,} \textbf{\bibinfo{volume}{249}},
  \bibinfo{pages}{L307} (\bibinfo{year}{1991}).

\bibitem[{\citenamefont{Fan and Ignatiev}(1990)}]{3xn5}
\bibinfo{author}{\bibfnamefont{W.~C.} \bibnamefont{Fan}} \bibnamefont{and}
  \bibinfo{author}{\bibfnamefont{A.}~\bibnamefont{Ignatiev}},
  \bibinfo{journal}{Phys. Rev. B.,} \textbf{\bibinfo{volume}{41}},
  \bibinfo{pages}{3592} (\bibinfo{year}{1990}).

\bibitem[{\citenamefont{Collazo-Davila
  et~al.}(1998)\citenamefont{Collazo-Davila, Grozea, and Marks}}]{HCC1}
\bibinfo{author}{\bibfnamefont{C.}~\bibnamefont{Collazo-Davila}},
  \bibinfo{author}{\bibfnamefont{D.}~\bibnamefont{Grozea}}, \bibnamefont{and}
  \bibinfo{author}{\bibfnamefont{L.~D.} \bibnamefont{Marks}},
  \bibinfo{journal}{Phys. Rev. Lett.,} \textbf{\bibinfo{volume}{80}},
  \bibinfo{pages}{1678} (\bibinfo{year}{1998}).

\bibitem[{\citenamefont{Lottermoser et~al.}(1998)\citenamefont{Lottermoser,
  Landemark, Smilgies, Nielsen, Feidenhans, Falkenberg, Johnson, Gierer,
  Seitsonen, Kleine, Bludau, Over, Kim and Jona}}]{HCC2}
  \bibinfo{author}{\bibfnamefont{L.}~\bibnamefont{Lottermoser}},
  \bibinfo{author}{\bibfnamefont{E.}~\bibnamefont{Landemark}},
  \bibinfo{author}{\bibfnamefont{D.-M.}~\bibnamefont{Smilgies}},
  \bibinfo{author}{\bibfnamefont{M.} \bibnamefont{Nielsen}},
  \bibinfo{author}{\bibfnamefont{R.}~\bibnamefont{Feidenhans}},
  \bibinfo{author}{\bibfnamefont{G.}~\bibnamefont{Falkenberg}},
  \bibinfo{author}{\bibfnamefont{R.~L.}~\bibnamefont{Johnson}},
  \bibinfo{author}{\bibfnamefont{M.}~\bibnamefont{Gierer}},
  \bibinfo{author}{\bibfnamefont{A.~P.}~\bibnamefont{Seitsonen}},
  \bibinfo{author}{\bibfnamefont{H.}~\bibnamefont{Kleine}},
  \bibinfo{author}{\bibfnamefont{H.}~\bibnamefont{Bludau}},
  \bibinfo{author}{\bibfnamefont{H.}~\bibnamefont{Over}},
  \bibinfo{author}{\bibfnamefont{S.~K.}~\bibnamefont{Kim}}, \bibnamefont{and}
  \bibinfo{author}{\bibfnamefont{F.}~\bibnamefont{Jona}}, \bibinfo{journal}{Phys. Rev. Lett.,}
  \textbf{\bibinfo{volume}{80}}, \bibinfo{pages}{3980} (\bibinfo{year}{1998}).

\bibitem[{\citenamefont{Erwin and Weitering}(1998)}]{HCC3}
\bibinfo{author}{\bibfnamefont{S.~C.} \bibnamefont{Erwin}} \bibnamefont{and}
  \bibinfo{author}{\bibfnamefont{H.~H.} \bibnamefont{Weitering}},
  \bibinfo{journal}{Phys. Rev. Lett.,} \textbf{\bibinfo{volume}{81}},
  \bibinfo{pages}{2296} (\bibinfo{year}{1998}).

\bibitem[{\citenamefont{Sch{\"a}fer et~al.}(2003)\citenamefont{Sch{\"a}fer,
  Erwin, Hansmann, Song, Rotenberg, Kevan, Hellberg, and Horn}}]{3xn6}
\bibinfo{author}{\bibfnamefont{J.}~\bibnamefont{Sch{\"a}fer}},
  \bibinfo{author}{\bibfnamefont{S.~C.} \bibnamefont{Erwin}},
  \bibinfo{author}{\bibfnamefont{M.}~\bibnamefont{Hansmann}},
  \bibinfo{author}{\bibfnamefont{Z.}~\bibnamefont{Song}},
  \bibinfo{author}{\bibfnamefont{E.}~\bibnamefont{Rotenberg}},
  \bibinfo{author}{\bibfnamefont{S.~D.} \bibnamefont{Kevan}},
  \bibinfo{author}{\bibfnamefont{C.~S.} \bibnamefont{Hellberg}},
  \bibnamefont{and} \bibinfo{author}{\bibfnamefont{K.}~\bibnamefont{Horn}},
  \bibinfo{journal}{Phys. Rev. B.,} \textbf{\bibinfo{volume}{67}},
  \bibinfo{pages}{85411} (\bibinfo{year}{2003}).

\bibitem[{\citenamefont{Over et~al.}(1994)\citenamefont{Over, Gierer, Bludau,
  Ertl, and Tong}}]{3xn7}
\bibinfo{author}{\bibfnamefont{H.}~\bibnamefont{Over}},
  \bibinfo{author}{\bibfnamefont{M.}~\bibnamefont{Gierer}},
  \bibinfo{author}{\bibfnamefont{H.}~\bibnamefont{Bludau}},
  \bibinfo{author}{\bibfnamefont{G.}~\bibnamefont{Ertl}}, \bibnamefont{and}
  \bibinfo{author}{\bibfnamefont{S.~Y.} \bibnamefont{Tong}},
  \bibinfo{journal}{Surf. Sci.,} \textbf{\bibinfo{volume}{314}},
  \bibinfo{pages}{243} (\bibinfo{year}{1994}).

\bibitem[{\citenamefont{de~Carvalho et~al.}(1984)\citenamefont{de~Carvalho,
  Cook, Cowell, Heavens, Prutton, and Tear}}]{Carvalho}
\bibinfo{author}{\bibfnamefont{V.~E.} \bibnamefont{de~Carvalho}},
  \bibinfo{author}{\bibfnamefont{M.~W.} \bibnamefont{Cook}},
  \bibinfo{author}{\bibfnamefont{P.~G.} \bibnamefont{Cowell}},
  \bibinfo{author}{\bibfnamefont{O.~S.} \bibnamefont{Heavens}},
  \bibinfo{author}{\bibfnamefont{M.}~\bibnamefont{Prutton}}, \bibnamefont{and}
  \bibinfo{author}{\bibfnamefont{S.~P.} \bibnamefont{Tear}},
  \bibinfo{journal}{Vacuum,} \textbf{\bibinfo{volume}{34}},
  \bibinfo{pages}{893} (\bibinfo{year}{1984}).

\bibitem[{\citenamefont{Palmino et~al.}(2005)\citenamefont{Palmino, Ehret,
  Mansour, Duverger, and Labrune}}]{Palmino}
\bibinfo{author}{\bibfnamefont{F.}~\bibnamefont{Palmino}},
  \bibinfo{author}{\bibfnamefont{E.}~\bibnamefont{Ehret}},
  \bibinfo{author}{\bibfnamefont{L.}~\bibnamefont{Mansour}},
  \bibinfo{author}{\bibfnamefont{E.}~\bibnamefont{Duverger}}, \bibnamefont{and}
  \bibinfo{author}{\bibfnamefont{J.-C.} \bibnamefont{Labrune}},
  \bibinfo{journal}{Surf. Sci.,} \textbf{\bibinfo{volume}{586}},
  \bibinfo{pages}{56} (\bibinfo{year}{2005}).

\bibitem[{\citenamefont{Pendry}(1980)}]{Rp}
\bibinfo{author}{\bibfnamefont{J.~B.} \bibnamefont{Pendry}},
  \bibinfo{journal}{J. Phys. C. Solid St. Phys.,}
  \textbf{\bibinfo{volume}{13}}, \bibinfo{pages}{937} (\bibinfo{year}{1980}).

\bibitem[{\citenamefont{Segall et~al.}(2002)\citenamefont{Segall, Lindan,
  Probert, Pickard, Hasnip, Clark, and Payne}}]{CASTEP}
\bibinfo{author}{\bibfnamefont{M.~D.} \bibnamefont{Segall}},
  \bibinfo{author}{\bibfnamefont{P.~J.~D.} \bibnamefont{Lindan}},
  \bibinfo{author}{\bibfnamefont{M.~J.} \bibnamefont{Probert}},
  \bibinfo{author}{\bibfnamefont{C.~J.} \bibnamefont{Pickard}},
  \bibinfo{author}{\bibfnamefont{P.~J.} \bibnamefont{Hasnip}},
  \bibinfo{author}{\bibfnamefont{S.~J.} \bibnamefont{Clark}}, \bibnamefont{and}
  \bibinfo{author}{\bibfnamefont{M.~C.} \bibnamefont{Payne}},
  \bibinfo{journal}{J. Phys.: Cond. Matt.,} \textbf{\bibinfo{volume}{14}},
  \bibinfo{pages}{2717} (\bibinfo{year}{2002}).

\bibitem[{\citenamefont{Probert and Payne}(2003)}]{Probert_Validation}
\bibinfo{author}{\bibfnamefont{M.~I.~J.} \bibnamefont{Probert}}
  \bibnamefont{and} \bibinfo{author}{\bibfnamefont{M.~C.} \bibnamefont{Payne}},
  \bibinfo{journal}{Phys. Rev. B.,} \textbf{\bibinfo{volume}{67}},
  \bibinfo{pages}{075204} (\bibinfo{year}{2003}).

\bibitem[{\citenamefont{Monkhorst and Pack.}(1976)}]{KP}
\bibinfo{author}{\bibfnamefont{H.~J.} \bibnamefont{Monkhorst}}
  \bibnamefont{and} \bibinfo{author}{\bibfnamefont{J.~D.} \bibnamefont{Pack}},
  \bibinfo{journal}{Phys. Rev. B.,} \textbf{\bibinfo{volume}{13}},
  \bibinfo{pages}{5188} (\bibinfo{year}{1976}).

\bibitem[{\citenamefont{Perdew et~al.}(1996)\citenamefont{Perdew, Burke, and
  Ernzerhof}}]{PBE}
\bibinfo{author}{\bibfnamefont{J.~P.} \bibnamefont{Perdew}},
  \bibinfo{author}{\bibfnamefont{K.}~\bibnamefont{Burke}}, \bibnamefont{and}
  \bibinfo{author}{\bibfnamefont{M.}~\bibnamefont{Ernzerhof}},
  \bibinfo{journal}{Phys. Rev. Lett.,} \textbf{\bibinfo{volume}{77}},
  \bibinfo{pages}{3865} (\bibinfo{year}{1996}).

\bibitem[{\citenamefont{Titterington and Kinniburgh}(1980)}]{CAVLEED}
\bibinfo{author}{\bibfnamefont{D.~J.} \bibnamefont{Titterington}}
  \bibnamefont{and} \bibinfo{author}{\bibfnamefont{C.~G.}
  \bibnamefont{Kinniburgh}}, \bibinfo{journal}{Comp. Phys. Comm.,}
  \textbf{\bibinfo{volume}{20}}, \bibinfo{pages}{237} (\bibinfo{year}{1980}).

\bibitem[{\citenamefont{Over et~al.}(1993)\citenamefont{Over, Huang, Tong, Fan,
  and Ignatiev}}]{Fan2}
\bibinfo{author}{\bibfnamefont{H.}~\bibnamefont{Over}},
  \bibinfo{author}{\bibfnamefont{H.}~\bibnamefont{Huang}},
  \bibinfo{author}{\bibfnamefont{S.~Y.} \bibnamefont{Tong}},
  \bibinfo{author}{\bibfnamefont{W.~C.} \bibnamefont{Fan}}, \bibnamefont{and}
  \bibinfo{author}{\bibfnamefont{A.}~\bibnamefont{Ignatiev}},
  \bibinfo{journal}{Phys. Rev. B.,} \textbf{\bibinfo{volume}{48}},
  \bibinfo{pages}{15353} (\bibinfo{year}{1993}).

\bibitem[{\citenamefont{Wigren et~al.}(1993{\natexlab{b}})\citenamefont{Wigren,
  Andersen, Nyholm, and Karlsson}}]{Wigren2}
\bibinfo{author}{\bibfnamefont{C.}~\bibnamefont{Wigren}},
  \bibinfo{author}{\bibfnamefont{J.~N.} \bibnamefont{Andersen}},
  \bibinfo{author}{\bibfnamefont{R.}~\bibnamefont{Nyholm}}, \bibnamefont{and}
  \bibinfo{author}{\bibfnamefont{U.~O.} \bibnamefont{Karlsson}},
  \bibinfo{journal}{Surf. Sci.,} \textbf{\bibinfo{volume}{293}},
  \bibinfo{pages}{254} (\bibinfo{year}{1993}{\natexlab{b}}).

\end{thebibliography}
\end{document}